\documentclass{mem}
\usepackage{natbib}\usepackage{txfonts}\usepackage{balance}
\usepackage{flushend}
\usepackage{graphicx}
\usepackage{xcolor}
\usepackage[breaklinks,pdftex]{hyperref}
\idline{75}{282}

\begin{document}
\def\teff{$T\rm_{eff }$}
\def\kms{$\mathrm {km s}^{-1}$}
\def\Ho{$H_0$}
\def\gaia{\textit{Gaia}}
\def\mWH{$m_H^W$}

\def\LCDM{$\Lambda$CDM}

\title{
Young stellar distance indicators and the extragalactic distance scale
}

\author{
Richard~I. Anderson
          }

\institute{
Institute of Physics, Laboratory of Astrophysics, \'Ecole Polytechnique F\'ed\'erale de Lausanne (EPFL), Observatoire de Sauverny, 1290 Versoix, Switzerland
}

\authorrunning{Anderson, R.~I.}

\titlerunning{Young stellar distance indicators}

\date{Received: 6 January 2023; Accepted: 19 January 2023}

\abstract{The extragalactic distance scale is perhaps the most important application of stellar distance indicators. Among these, classical Cepheids are high-accuracy standard candles that support a $1.4\%$ measurement of Hubble's constant, $H_0$. The accuracy of Cepheid distances is thus directly relevant for understanding the implications of the {\it Hubble tension}, the $>5\sigma$ discord among direct, late-Universe $H_0$ measurements and $H_0$ values inferred from the early Universe observations assuming $\Lambda$CDM cosmology. This invited review aims to provide an accessible overview of the state of the art distance ladder that has established the Hubble tension, with a focus on Cepheids, their absolute calibration using trigonometric parallaxes from the ESA mission {\it Gaia}, and other Cepheid-related systematics. New observational facilities such as {\it JWST} and upcoming large surveys will provide exciting avenues to further improve distance estimates based on Cepheids.

}
\maketitle{}

\section{Cepheids and Leavitt's law}
Classical Cepheids are evolved intermediate-mass stars primarily observed during the blue loop phase of stellar evolution that occurs while the core is burning helium. A rare, but very interesting, subset of classical Cepheids is observed prior to core-He ignition during the Hertzsprung gap (the first instability strip crossing). It is also crucial to distinguish between classical (type-I) Cepheids and type-II Cepheids \citep{Baade1956}, whose sub-types include the W~Virginis, BL~Herculis, and RV~Tauri stars \citep[e.g.][and references therein]{Jurkovic2021}. While they share a similar name and similar photometric variability features and timescales, type-II Cepheids are several Gyr old evolved low-mass stars that are physically very different from type-I Cepheids. In the following, we use the term ``Cepheids'' to refer to the objects most relevant for the distance ladder, that is, core He burning (second or third crossing) classical Cepheids that pulsate in the fundamental mode unless explicitly stated otherwise, whose prototype is $\delta$~Cephei. Although Cepheids are by no means the only young stellar distance indicators, their unique importance for the extragalactic distance scale hopefully justifies the omission of other young stellar distance indicators \citep[such as A\&B-type supergiants or early-type eclipsing binaries, cf.][]{Kudritzki2003,Taormina2019}.

Concerning their ``youth'', Cepheids of Solar metallicity are of order $30-300$\,Myr old, and age estimates based on model-dependent period-age relations depend strongly on mixing processes (notably rotation) affecting the Main Sequence evolution of the progenitors \citep{Bono2005,Anderson2016rot}. Chemical composition also plays an important role by allowing for older, lower-mass Cepheids at lower metallicity and rendering Cepheid ages slightly older at identical pulsation period.

Cepheids were the first pulsating stars discovered to obey a linear relation between the logarithm of their variability period, $\log{P}$, and their intrinsic brightness based on 25 stars in the Small Magellanic Cloud \citep{Leavitt1908,Leavitt1912}. The aptly named Leavitt law (henceforth: LL) thus relates absolute magnitude $M$ to $\log{P}$, so that distance modulus $\mu = m - M$ can be calculated from apparent magnitudes, $m$, and variability periods $P$ once the LL is calibrated. \citet{Hertzsprung1913} quickly realized the importance of this relation for distance estimation. Over time, the LL has allowed great progress in the understanding of the size of the Milky Way, the Local Universe, and its expansion rate. 

The Cepheid LL is usually calibrated using the form:
\begin{equation}
M = \alpha + \beta \cdot \log{P/P_0} + \gamma \cdot \mathrm{[Fe/H]} \ ,
\label{eq:LL}
\end{equation}
with $\alpha$ the fiducial absolute magnitude of a Solar metallicity Cepheid at the pivot period $P_0$, $\beta$ the LL slope, and $\gamma$ the metallicity term that allows to correct the impact of chemical composition using iron (also: oxygen) abundances. Different LL variants have been adopted, including broken (at $P_0$) slopes \citep{Riess2016,Bhardwaj2016}, and models indicate a possible LL slope-dependence on metallicity \citep{Anderson2016rot}. 

LLs are calibrated using Cepheids whose distance $d$ is either measured individually or known as a sample average using $M = m - \mu = m - 5 \log{d} + 5$. Trigonometric parallaxes are today's Gold standard for Milky Way Cepheids \citep{Feast1997,Benedict2007,Casertano2016,Riess2018plx,CruzReyes22}, in particular thanks to the unprecedented number and quality of parallaxes published by the ESA mission {\it Gaia} \citep[e.g.][and accompanying documentation]{Prusti2016gaia,Brown2021gaia}, cf. Sect.\,\ref{sec:plx} and Clementini (this volume). Baade-Wesselink-type distances of Cepheids have not been favored in recent analyses due to systematic uncertainties related to the projection factor, cf. Nardetto (this volume). 

Sample average distances can be used to calibrate Cepheids observed in other galaxies whose intrinsic depth is insignificant compared to their distance from the Sun. An important exception is the Large Magellanic Cloud (LMC), which is host to several thousand Cepheids of various pulsation modes \citep{Soszynski2019}, and whose distance has been measured to an exquisite relative uncertainty of $1\%$ using detached eclipsing red giant binary systems \citep[DEBs][]{Pietrzynski2019}. At this level of precision, the line-of-sight depth of the LMC exceeds the intrinsic dispersion of the LL at LMC metallicity, so that corrections to the mean distance of the LMC DEBs are required to minimize the observed LL scatter and obtain the most accurate calibration \citep[e.g.][]{Breuval22}. In practice, this correction corresponds to a small magnitude offset calculated using the on-sky distance of Cepheids from the  LMC's major axis determined using the 20 DEBs. An analogous and larger depth effect  complicates LL calibration using DEB distances of the SMC \citep{Graczyk2020}, whose major axis is nearly aligned with the line of sight \citep[e.g.][]{Scowcroft2016}. At distances greater than a few hundred kpc, no further ``geometric corrections'' are required at the current level of precision. Cepheids in the most nearby spiral galaxies, such as Andromeda (M31) and M33, provide useful cross-checks \citep[e.g.][]{Li2021,Pellerin2011}, and {\it Gaia} DR3 recently delivered time-series photometry to Cepheids in both galaxies \citep{GAPS}. Other nearby galaxies are particularly useful for understanding metallicity differences \citep[e.g.][]{Bernard2013} and the consistency of distance estimates based on different stellar standard candles \citep[e.g.][]{Lee2022}. 

The spiral galaxy NGC\,4258 (M106) is the most distant galaxy that contributes directly to absolute LL calibration thanks to water MASERs orbiting its central supermassive black hole. Modeling MASER features tracked over the course of many years using a warped Keplerian disk model has resulted in a $1.4\%$ geometric distance \citep{Reid2019}. An important benefit of using NGC\,4258 for LL calibration is that its Cepheids are observed in a very similar context as even more distant Cepheids, notably with respect to the camera setup, signal-to-noise, and crowding properties, among other things \citep{Yuan2022}.

\section{Trigonometric parallaxes\label{sec:plx}}
Prior to the first \gaia\ Data Release in 2016, parallaxes of Cepheids precise to better than $\sim 10\%$ were very rare. The ESA mission {\it Hipparcos} had provided parallaxes for $24$ MW Cepheids, which \citet{Feast1997} used to calibrate the LL and estimate the distance to the LMC. However, the Pleiades highlighted a potential problem involving {\it Hipparcos} parallaxes, prompting \citet{Benedict2007} to measure narrow-angle parallaxes of 10 MW Cepheids using the {\it HST} Fine Guidance Sensor. A re-reduction of {\it Hipparcos} parallaxes by \citet{Hipparcos2007,VanLeeuwen2007} showed promise for improvement. Yet, the Pleiades problem persisted and long baseline radio interferometry \citep{Melis2014} showed conclusively that {\it Hipparcos} parallaxes were wrong at least in some cases. \citet{Riess2014,Casertano2016,Riess2018plx} used the spatial scanning mode of {\it HST}/WFC3 to measure $10\%$ parallaxes at distances up to a few kpc, notably including a significant new number of long-period Cepheids for the MW sample. The importance of \gaia\ for Cepheid parallaxes was immediately clear from \gaia\ DR1  \citep{Lindegren2016,Clementini2017gaia}. Clementini (this volume) presents an overview of improvements across the various \gaia\ data releases. 

\gaia\ parallaxes are the undisputed Gold standard for DL calibration. However, there unfortunately has remained a parallax bias issue discovered via an average non-zero parallax value of quasars. Using several thousand bright stars, of order $10^5$ stars in the LMC, and millions of quasars, \citet{Lindegren2021} have understood this parallax bias to correlate with several features and provided a correction. In short, the parallax bias depends on the sine of the ecliptic latitude (an artefact of \gaia's scanning law?), apparent \gaia\ $G-$band magnitude (an artefact of \gaia's complex photometric processing that involves a gating mechanism to avoid saturation as well as differences in image treatment involving either 2D or 1D point/line spread functions determined using magnitude-dependent window sizes \citep{Riello2021,Lindegren2021}, and source color (possibly related to chromatic aberration). The recipe for bias correction works very well at magnitudes fainter than $\approx 11-12$\,mag, where a sufficient number of objects was available to correct these correlations \citep[e.g.][]{Zinn2021,MaizApellaniz2022}. However,  geometric distance measurements  of brighter stars have revealed residual parallax offsets in stars $G \lesssim 10$\,mag \citep[e.g.][]{riess21,Zinn2021}. 

Recently, \citet{riess2022cluster} and \citet{CruzReyes22} used MW Cepheids residing in open clusters to resolve the problem of the residual parallax offset, demonstrating multiple significant advantages of cluster Cepheids. First, the large number of cluster member stars improves statistical precision on the cluster's average parallax, resulting in statistical uncertainties as low as $1.4\,\mu$as. Second, the parallax bias correction is accurate in the color and magnitude range spanned by cluster member stars. The dominant uncertainty of cluster parallaxes is thus set by the angular covariance of \gaia\ parallaxes \citep{Lindegren2021}. Accounting for angular covariance and underestimated uncertainties \citep{MaizApellaniz2022} raises the typical total parallax uncertainty of cluster Cephieds to $7\,\mu$as, or $1/3$ that of field Cepheids. Since luminosity depends on distance squared, a single cluster Cepheid contributes as much as 9 field Cepheids to LL calibration. \citet{CruzReyes22} thus calibrated the absolute magnitude of a 10\,d Cepheid to $0.9\%$ in two independent photometric data sets from \gaia- and {\it HST}, while measuring the residual parallax offset of field Cepheids by comparing their parallaxes to cluster parallaxes.

\section{The Distance ladder and $H_0$\label{sec:H0}}
Hubble's constant, \Ho, quantifies the expansion rate of the Universe today (at $z=0$) and is one of the most important parameters for cosmology and extragalactic astronomy because it sets the size of the observable Universe and tells us its age. The Hubble-Lema\^itre law, $H_0 = v/D$, relates (apparent) recession velocities $v\approx cz$ ($v \ll c$) to luminosity distance $D_L$. In reality, $v$ is not a \textit{velocity}, but rather a redshift caused by cosmic expansion. The distinction is important, since velocities are confined to $v < c$, whereas $cz$ can readily exceed $c$. More generally, $H_0$ is related to luminosity distance via the Friedmann equation and the Robertson-Walker metric (expanded here to second order):
\begin{equation}
D_L = \frac{cz}{H_0} \left[ 1 + \frac{1}{2}(1-q_0)z - O(z^2) +  \ldots \right] \ ,
\end{equation}
where $q_0 \approx -0.55$ is the \textit{deceleration} parameter, whose observed negative value implies the Universe's accelerated expansion \citep{riess1998,Perlmutter1999}. 

\Ho\ is best measured at small non-zero redshifts (in the Hubble flow), where space is expanding isotropically, acceleration is not significant, and peculiar motions are subdominant, roughly at distances of $90 - 600$\,Mpc \citep[$0.01 < z < 0.0233$]{Riess2016}. Mapping such distances requires intrinsically extremely bright objects, and type-Ia supernovae (SNeIa) are well-suited to this end thanks to extreme luminosity ($M_B \approx -19.25$\,mag) and their ability to provide $\sim 3\%$ relative distances per SNIa. However, SNeIa are very rare, and no SNIa has as yet been observed in a galaxy with a precisely and directly measured distance. SNeIa thus require external calibration by other means, such as Cepheids. 

Conceptually, the distance ladder (DL) consists of three rungs. The first rung calibrates the Cepheid LL using direct distances, the second rung calibrates the SNeIa absolute magnitude using Cepheids (out to currently $\sim 70$\,Mpc), and the last rung maps the Hubble flow using SNeIa. However, today's modern DL is built using principles of ``ladder safety'' to prevent hazardous missteps and tightly links all rungs together in a global least squares fit. Thus, the DL internally benefits from great statistical precision, while its accuracy is largely dependent on the external absolute calibration. In its matrix formulation, the DL uses the data vector $y$, the design matrix $L$ that encodes the relevant equations, such as the LL, distance moduli, etc., and the covariance matrix $C$ to determine the best-fit parameter vector $q$ by minimizing: 
\begin{equation}
\chi^2 = (y - Lq)^T C^{-1} (y - Lq)  \ .
\label{eq:DL}
\end{equation}
In the latest SH0ES distance ladder, a very significant effort has been made to quantify and include off-diagonal elements in the covariance matrix, such as correlated background noise \citep{riess22}. Using 3445 degrees of freedom, this least squares procedure yields 5 best fit parameters: the LL slope $\beta$ ($b$), metallicity effect $\gamma$ ($Z_W$), the fiducial absolute magnitude of a 10-day Cepheid in the {\it HST} NIR Wesenheit magnitude $M_W^0$, the absolute $B-$band magnitude of SNeIa, $M_B^0$, and Hubble's constant $5 \log_{10}{H_0}$. Alternative symbols from the literature are listed in parenthesis. The matrix formalism is both simple and accurate, and allows to readily re-determine $H_0$ for different analysis variants, such as broken LL slopes,  reddening laws, data selections, etc. However, the results from the matrix formalism are cross-checked with computationally intensive Markov chain Monte Carlo (MCMC) simulations that allow inspection of marginalized posterior distributions and correlations among fit parameters. TRGB distances have also been included in this procedure and can help to further improve precision. Including the latest results based on cluster Cepheids, this DL yields $H_0= 73.15 \pm 0.97 \, \mathrm{km\,s^{-1}\,Mpc^{-1}}$ \citep{riess2022cluster}.

Seen the other way around, the DL links the distance-redshift relation of SNeIa to the absolute angular scale provided by {\it Gaia}'s trigonometric parallax measurements. This is particularly powerful, since the Hubble diagram can also be connected to the early Universe's angular scale, the cosmological sound horizon, using an inverse distance ladder \citep{Lemos2019}. Hence, the DL allows to connect angular scales at opposite ends of the cosmos to each other\hbox{---}trigonometric parallaxes and the quantum fluctuations of the early Universe\hbox{---}and provides a crucial end-to-end test of observational cosmology \citep{riess20nature}.

\section{The Hubble tension and focus on systematics}
Therefore, if the $\Lambda$ Cold Dark Matter (\LCDM) concordance cosmological model provides an accurate representation of the physics governing the Universe and its evolution, one would expect local \Ho\ measurements to match values of \Ho\ inferred from early Universe observations, such as the Cosmic Microwave Background. However, a $\sim 5\sigma$ disagreement between late and early Universe $H_0$ has appeared since 2015 and is often referred to as the {\it Hubble tension} \citep{Riess2016,riess22}. If the tension's origin can be firmly attributed to the early Universe \Ho\ determination, then it would be likely for \LCDM\ to be incomplete, requiring revision by as yet unknown physics \citep[for possible options, cf.][]{snowmass2021}. However, such extraordinary claims require extraordinary evidence, so much effort is under way to further improve the accuracy of the DL-based $H_0$. In this process, it remains important to lower statistical uncertainties to measure \Ho\ to a precision similar to that of the early-Universe value, that is, to better than $1\%$. In the process, the focus of recent work has increasingly turned to systematics.

The term `systematics' encompasses both a) uncertainties that cannot be improved with larger samples and b) biases that systematically shift measurements away from the truth. 
Strategies for mitigating systematics include (in arbitrary order) a) {\bf favoring data insensitive to specific biases} (e.g., infrared data minimize uncertainties related to extinction), b) {\bf maximizing data homogeneity} to avoid errors due to transformations (e.g., exclusive use of {\it HST} photometry), c) ensuring the {\bf physical similarity} of objects along the distance ladder (e.g., matching period ranges of Cepheids near and far, correcting metallicity differences), d) simultaneously fitting all data including covariance information, e) {\bf accounting for differences introduced by the observational setup} (e.g., CRNL corrections or stellar association bias), and f) incorporating {\bf corrections for other physical effects} (e.g., relativistic effects), among other things. 

Exclusive use of the {\it HST} photometric system has been made possible by two recent non-standard observing modes, the drift scanning mode applied to MW Cepheids \citep{riess2018phot,riess21}, and the DASH mode for Cepheids in the Magellanic Clouds \citep{riess2019}. Count-rate non-linearity (CRNL) must be applied to WFC3/IR data when comparing Cepheids across a dynamic range of $16$\,mag, or a flux ratio of $2.5$ million.

Extinction is effectively mitigated by the Wesenheit\footnote{The German word ``Wesenheit'' relates to the abstract innate nature (or essence) of an entity. If you are having difficulty with this word, fear not: ``Wesenheit'' was always a high-brow word used by few, and its use has dwindled even more since the 1970s, cf. \url{https://www.dwds.de/wb/Wesenheit}.} function \citep{VanDenBergh1975,Madore1982}. Wesenheit magnitudes, $m^W$, are constructed to be reddening-free assuming a specific reddening law \citep[SH0ES uses $R_V=3.3$ from][]{Fitzpatrick1999,Schlafly2011}. For example, the SH0ES near-IR Wesenheit formalism combines intrinsically extinction-insensitive $H-$band (F160W) magnitudes with a small offset based on optical color,  \mWH$= \mathrm{F160W} - R^W\cdot(\mathrm{F555W}-\mathrm{F814W})$. Here, $R^W = A_H / (A_V - A_I) \approx 0.4$ is given by the reddening law. For Gaia, $R^W_G = A_G / (A_{Bp} - A_{Rp}) \approx 1.91$. The use of wide-band photometry for the distance scale implies that $R^W$ depends on a star's intrinsic color, since the shape of the spectral energy distribution incident on the photometric passband should not be neglected \citep{anderson22}. A (unlikely) systematic due to reddening could arise if there were a systematic difference between the thousands of sightlines among Cepheids in anchor galaxies (MW, LMC, NGC4258) and the thousands of sightlines to Cepheids in SN-host galaxies. \citet{Moertsell2022,riess22} recently discussed this effect and concluded that extinction cannot explain the Hubble tension.

The physical similarity of Cepheids in anchor and SN-host galaxies is being ensured by considering their characteristic light curves. Extragalactic Cepheid candidates pass several selection criteria concerning their mean color, amplitude ratios in $V$ and $I-$band (where available), and distance from the galaxy's LL by $\sigma-$clipping. The observed similarity of the Hertzsprung progression across the DL rungs is very strong evidence that Cepheid samples are drawn from a common population. 

Corrections for the effect of metallicity on Cepheid luminosity have recently made significant progress, benefiting from a reanalysis of spectroscopic abundances of LMC and MW Cepheids \citep{romaniello22,Ripepi2022,Trentin22} and improved accuracy of \gaia\ parallaxes. In particular, \citet{Breuval22} recently calibrated $\gamma$ using the metallicity range spanned by MW, LMC, and SMC Cepheids and found results consistent with the metallicity term used in the SH0ES DL (labeled $Z_W$ in SH0ES). However, as pointed out by \citet{riess22}, the metallicity range of Cepheids in SN-host and anchor galaxies is comfortably contained between MW and LMC Cepheids. Hence, the characterization over a longer metallicity lever allows the accurate correction of metallicity effects in the more restricted range. 

Crowding corrections are required to remove unwanted light contributions due to blending, that is, the statistical superposition of stars \citep[e.g.][]{Yuan2022jwst}. Although crowding corrections accurately and statistically without bias remove light contributions, they impose a penalty in terms of precision, and it would be preferable to avoid blending in the first place. Thankfully, the much improved spatial resolution of {\it JWST} ($0.03"$/pixel vs $0.13"$/pixel in {\it HST} WFC3/IR) will very soon enable significant precision gains by instrumentally ``uncrowding'' Cepheids in SN host galaxies \citep{Yuan2022jwst}. 

Stellar association bias differs from other crowding in that it is incurred due to a physical association of stars rather than chance blending. Moreover, this bias is not removed by crowding corrections due to limited spatial resolution \citep{anderson18}. In short, the detector's finite angular resolution corresponds to increasing physical sizes as a function of distance. Cepheids occurring in or near their birth clusters thus increasingly blend with their host clusters with distance. For a typical cluster size of $4\,pc$, cluster Cepheids are essentially unresolved at distances of $> 10$\,Mpc. Crowding corrections based on field stars probe larger angular scales, and therefore are insensitive to the light contributed by host clusters. In the MW and the LMC, Cepheids are well-resolved from their host clusters. Thus the light contribution from host clusters is not counted in most Cepheids that calibrate the LL, while it is present for more distant Cepheids. \citet{anderson18} estimated this effect to be of the order of $0.3\%$ on \Ho\ using M31 as a SN-host proxy, and \citet{riess22} has since applied a correction of $0.07$\,mag to extragalactic Cepheids to mitigate the effect. Spetsieri et al. (in prep.) presented during the meeting our ongoing work to quantify stellar association bias using {\it HST} UV observations of the closest SN-host galaxy M101.

Relativistic corrections are required to account for the small, albeit one-sided, systematic difference between the inertial frames where the Cepheid LL is calibrated and where it is applied. \citet{anderson19} pointed out that time-dilation causes distant ($z > 0$) Cepheid periods to appear longer by a factor $\Delta \log{P} = \log{(1+z)}$ compared to their rest-frame pulsation periods. Correcting this bias interestingly increased \Ho\ and strengthened the Hubble tension. Dilation of Cepheid periods has since been incorporated into the SH0ES DL \citep{riess22} and contributes systematically to the $5.0\sigma$ Hubble discord. Further relativistic corrections were recently explored, including $K-$corrections and the effect of redshift on Wesenheit magnitudes \citep{anderson22}. Interestingly, $K-$corrections applicable to \mWH\ are negligible because $K-$corrections to $H-$band magnitudes and $0.4\times(V-I)$ nearly compensate. However, single-band {\it JWST} observations of Cepheids or stars near the Tip of the Red Giant Branch at $100$\,Mpc require $K-$corrections of order $1\%$.

\section{Conclusions}

Classical Cepheids are excellent standard candles that have enabled a $1.4\%$ measurement of the Hubble constant \citep{riess22,riess2022cluster}. One of their key strengths is that every Cepheid is a uniquely identifiable standard candle. This allows to directly measure a host of properties and cleanly investigate systematics, such as the effect of metallicity. One of the current limitations to the precision of the distance ladder is blending, which will be significantly improved thanks to {\it JWST}'s $\sim 4\times$ better spatial resolution. Preliminary analysis of a serendipitous (non-Cepheid targeting) observation of NGC\,1365 nicely illustrates {\it JWST}'s potential \citep{Yuan2022jwst}. {\it Gaia} is expected to further improve the absolute LL calibration, both by improved parallaxes of Cepheids and a greater number of cluster Cepheids in future data releases \citep[e.g.][]{CruzReyes22}. Besides their ability to calibrate the DL, Cepheids will also continue to play a vital role in understanding the size and structure of the Milky Way (cf. Grebel, this volume) and the local Universe in general.  

Large surveys, both photometric (notably Rubin/LSST) and spectroscopic (4MOST, WEAVE, etc.), will yield a wealth of relevant information that will help to both better understand our most precise stellar standard candles and to further understand and mitigate systematics affecting distances. In particular the ability to combine multi-band time-series data should lead to new insights, as will the study of long and precise time-series data with asteroseismic potential \citep{sueveges18,anderson20,Smolec2022}. Last, but not least, the advent of 30m-class telescopes, such as ESO's ELT, will allow to obtain much more detailed observations of extragalactic Cepheids, by even further improving spatial resolution at unprecedented distances and enabling spectroscopy of individual stars beyond the Magellanic Clouds. This wealth of upcoming opportunities makes young stellar distance tracers a timely and exciting subject to investigate, and we can be sure to be surprised in the future.

\begin{acknowledgements}
I congratulate the organizers for handling the significant weather-related obstacles so successfully and for a wonderful conference. I acknowledge support from the European Research Council (ERC) under the European Union's Horizon 2020 research and innovation programme (Grant Agreement No. 947660) and through a Swiss National Science Foundation Eccellenza Professorial Fellowship (award PCEFP2\_194638).
\end{acknowledgements}

\bibliographystyle{aa}
\bibliography{bibliography.bib}

\end{document}